\documentclass[
reprint, 
aps, prl, superscriptaddress, 
nofootinbib, 
bibnotes,
amsmath,amssymb,
floatfix,
]{revtex4-2}

\usepackage{graphicx}

\usepackage{natbib}
\usepackage[colorlinks=true]{hyperref}

\usepackage{braket}
\usepackage[utf8]{inputenc}
\usepackage{subfigure}

\newcounter{subeqn}

\usepackage{setspace}
\usepackage{color}
\usepackage{float}
\usepackage[english]{babel}
\usepackage{enumitem}
\usepackage{placeins}
\usepackage[normalem]{ulem}

\renewcommand*{\thefootnote}{\fnsymbol{footnote}}

\begin{document}
\title {Thorium-229 in its Highest Charge States: Single-Ion Nuclear Clocks for Tests of Fundamental Interactions}
\author{C.~Brandau}
\thanks{These authors contributed equally to this work}
\affiliation{GSI Helmholtzzentrum f\"ur Schwerionenforschung, Planckstra\ss e 1, D-64291 Darmstadt, Germany}

\author{P.~Micke}
\thanks{These authors contributed equally to this work} 
\affiliation{GSI Helmholtzzentrum f\"ur Schwerionenforschung, Planckstra\ss e 1, D-64291 Darmstadt, Germany}
\affiliation{Helmholtz-Institut Jena, Fr\"obelstieg 3, D-07743 Jena, Germany}
\affiliation{Institut f\"ur Optik und Quantenelektronik, Friedrich-Schiller-Universit\"at Jena, Max-Wien-Platz 1, D-07743 Jena, Germany}
\affiliation{Abbe Center of Photonics, Albert-Einstein-Stra\ss e 6, D-07745 Jena, Germany}

\author{I.~Hartl}
\affiliation{Deutsche Elektronen-Synchrotron DESY, Notkestra\ss e 85, D-22607 Hamburg, Germany}

\author{C.~M.~Heyl}
\affiliation{GSI Helmholtzzentrum f\"ur Schwerionenforschung, Planckstra\ss e 1, D-64291 Darmstadt, Germany}
\affiliation{Helmholtz-Institut Jena, Fr\"obelstieg 3, D-07743 Jena, Germany}
\affiliation{Deutsche Elektronen-Synchrotron DESY, Notkestra\ss e 85, D-22607 Hamburg, Germany}

\author{P.~Indelicato}
\affiliation{Laboratoire Kastler Brossel, Sorbonne Université, CNRS, ENS-PSL Research University, Collège de France, Case 74, 4 place Jussieu, F-75005 Paris, France}

\author{Yu.~A.~Litvinov}
\affiliation{GSI Helmholtzzentrum f\"ur Schwerionenforschung, Planckstra\ss e 1, D-64291 Darmstadt, Germany}
\affiliation{Institut f\"ur Kernphysik, Universit\"at zu K\"oln,  Z\"ulpicher Stra\ss e 77, D-50937 K\"oln, Germany }

\author{R.~X.~Schüssler}
\affiliation{GSI Helmholtzzentrum f\"ur Schwerionenforschung, Planckstra\ss e 1, D-64291 Darmstadt, Germany}
\affiliation{Helmholtz-Institut Jena, Fr\"obelstieg 3, D-07743 Jena, Germany}

\author{M.~Vogel}
\affiliation{GSI Helmholtzzentrum f\"ur Schwerionenforschung, Planckstra\ss e 1, D-64291 Darmstadt, Germany}

\author{Th.~St\"ohlker}
\affiliation{GSI Helmholtzzentrum f\"ur Schwerionenforschung, Planckstra\ss e 1, D-64291 Darmstadt, Germany}
\affiliation{Helmholtz-Institut Jena, Fr\"obelstieg 3, D-07743 Jena, Germany}
\affiliation{Institut f\"ur Optik und Quantenelektronik, Friedrich-Schiller-Universit\"at Jena, Max-Wien-Platz 1, D-07743 Jena, Germany}
\affiliation{Abbe Center of Photonics, Albert-Einstein-Stra\ss e 6, D-07745 Jena, Germany}
\date{\today}

\begin{abstract}
The prospects and the implementation of single-ion nuclear clocks of $^{229}$Th$^{q+}$ ions in their highest charge states $q=90, \ldots, 87$ are discussed.  Highly-ionized-thorium clocks are ideal for tests of fundamental interactions since the ions are elementary quantum systems composed of only a few building blocks. Two cases of $^{229}$Th$^{q+}$ clocks excel: a) one-electron $^{229}$Th$^{89+}$ that combines two nuclear-clock transitions in the VUV with hyperfine IR atomic-clock transitions, and, b) fully ionized $^{229}$Th$^{90+}$ which constitutes the prototype of a nuclear clock, one without any electrons. We evaluate the feasibility of such clocks by means of quantum logic spectroscopy (QLS) in linear Paul traps. Due to its universal nature, the QLS approach allows for systematic clock comparisons using different charge states as well as different spectroscopy transitions on the same experimental platform. A valuable asset towards single-ion $^{229}$Th$^{q+}$ clocks is the process of nuclear hyperfine mixing that enables the tunability of the natural linewidth of the clock transition over more than five orders of magnitude by changing the charge state. 
\end{abstract}

\pacs{xx.80.Lx, xx.10.+z, xx.30.J-}

\maketitle

\renewcommand{\thefootnote}{\arabic{footnote}}
 \setcounter{footnote}{0}

About half a century after initial hints on a nuclear-level doublet in $^{229}$Th that is split by an exceptionally low energy separation of a few eV \cite{Kroger1976:FeaturesLowenergyLevel_,Reich1990:EnergySeparationDouble_}, in 2024, three different groups published the laser excitation of this transition \cite{Tiedau2024:LaserExcitationTh229N_,Elwell2024:LaserExcitation$^229ma_,Zhang2024:FrequencyRatio229mThN_}.
In the latter experiment \cite{Zhang2024:FrequencyRatio229mThN_}, by comparison to a $^{87}$Sr atomic clock, the isomeric $(m)$
transition energy of thorium nuclei embedded in a CaF$_2$ host crystal, effectively $^{229}$Th$^{4+}$, was measured to $\nu_\gamma^{(m)} = 2 020.407 384 335(2)$~THz, 
correspoding to about $E_\gamma^{(m)} = 8.355 733 55$~eV  or a vacuum wavelength of $148.382 182$ nm. These initial laser experiments are recognized as a major breakthrough for the development of a nuclear clock, i.e., a timekeeper based on a nuclear transition instead of an atomic transition \cite{Palffy2024:CountdownNuclearClock_,Gibney2024:NuclearClockBreakthrou_}. The idea of a nuclear clock was developed by Peik and Tamm more than 20 years ago \cite{Peik2003:NuclearLaserSpectrosco_, Peik2009:ProspectsNuclearOptica_}. They also emphasized a main advantage of a nuclear clock: the insensitivity of the nuclear transition to external perturbations from electric or magnetic fields. A nuclear clock promises uncertainties at the 19$^\text{th}$ digit and beyond \cite{Peik2003:NuclearLaserSpectrosco_,Peik2009:ProspectsNuclearOptica_, Campbell2012:SingleIonNuclearClock_} and potentially competes with or even excels present optical atomic clocks that likewise nowadays reach $10^{-19}$ uncertainty \cite{Filzinger2026:MultiionOpticalClock$_,Marshall2025:HighStabilitySingleIon_}.

$^{229}$Th nuclear clocks shine as unique tools for tests of fundamental physics, namely, to query temporal variations of fundamental constants, as unique sensors for ultra-light dark matter or for the quest of a potential fifth fundamental force \cite{Peik2021:NuclearClocksTestingF_, Thirolf2024:ThoriumIsomer$$^229m$$_,Safronova2019:SearchVariationFundame_,Flambaum2006:EnhancedEffectTemporal_,Fadeev2020:Sensitivity$^229mathrmT_, Fadeev2022:EffectsVariationFineS_, Berengut2009:ProposedExperimentalMe_, Beeks2025:FinestructureConstantS_,Caputo2025:SensitivityNuclearCloc_, Dzuba2025:UsingThIiiIon_,Fuchs2025:SearchingDarkMatter$^_,Zaheer2025:QuantumMetrologyAlgori_,Delaunay2025:ProbingNewForcesNucle_}:
The high sensitivity to fundamental  interactions stems from an almost complete cancellation of large nuclear potentials of ground and isomeric state \cite{Flambaum2006:EnhancedEffectTemporal_} such that small variations of the 
coupling  of quantum electrodynamics (QED) or of quantum chronodynamics (QCD) would sensitively enter the isomer transition energy or change the measured lineshape of the transition \cite{Fuchs2025:SearchingDarkMatter$^_}. The according sensitivities are commonly parameterized by the enhancement factors  $K_\alpha$ (QED) and $K_q$ (QCD). These factors quantify the change of a transition frequency $\delta f / f = K_\alpha \cdot \delta \alpha / \alpha$ and $\delta f / f = K_q \cdot \delta m_q / m_q $ under variation of the fine-structure constant $\alpha$ or the dimensionless parameter $X_q = m_q / \lambda_\mathrm{QCD}$, respectively. In the latter,  $m_q = m_u + m_d /2$ is the average of the up and down quark-masses and $\lambda_\mathrm{QCD}$ the QCD scale \cite{Flambaum2006:EnhancedEffectTemporal_,Fadeev2020:Sensitivity$^229mathrmT_,Fadeev2022:EffectsVariationFineS_}. Typically, in calculations of $K_q$, $\lambda_\mathrm{QCD}$ is kept constant. Actual values of $K_\alpha$ and $K_q$ for the isomeric transition of $^{229}$Th can be deduced from measured nuclear parameters  \cite{Flambaum2006:EnhancedEffectTemporal_, Fadeev2022:EffectsVariationFineS_, He2008:TemporalVariationFine_,Berengut2009:ProposedExperimentalMe_,Fadeev2020:Sensitivity$^229mathrmT_, Beeks2025:FinestructureConstantS_,Caputo2025:SensitivityNuclearCloc_}. Using experimental data from \cite{Safronova2018:NuclearChargeRadii$^2_,Yamaguchi2024:LaserSpectroscopyTripl_,Zhang2024:FrequencyRatio229mThN_}, for $^{229}$Th, $K_q \approx K_\alpha \approx 10^4$ \cite{Beeks2025:FinestructureConstantS_,Caputo2025:SensitivityNuclearCloc_,Fuchs2025:SearchingDarkMatter$^_} is derived. A $^{229}$Th-clock is thus about 3 to 4 orders of magnitude more sensitive to time-dependent variations of QED or QCD than present optical atomic clocks \cite{Fadeev2022:EffectsVariationFineS_}. Similarly, besides $^{235}$U, in which $K_\alpha \approx  K_q \approx 1000 $, the sensitivities of other Mössbauer nuclear transitions are $K_\alpha \sim \mathcal{O}(1)$  and $K_q \sim \mathcal{O}(1)$ \cite{Fadeev2022:EffectsVariationFineS_}. 

Existing thorium-clock concepts focus on low charge states of $^{229}$Th$^{q+}$ up to $q=4$: Either $^{229}$Th$^{4+}$ embedded in the lattice of host crystals  \cite{Tiedau2024:LaserExcitationTh229N_,Elwell2024:LaserExcitation$^229ma_,Zhang2024:FrequencyRatio229mThN_},
using thin films of $^{229}$Th-compounds in a conversion electron Mössbauer spectroscopy (CEMS) arrangement \cite{vonderWense2020:$$^229$$ThIsomerProspe_,Elwell2025:LaserbasedConversionEl_}, or lowly charged $^{229}$Th$^{q+}$ ions in Paul traps are considered \cite{Peik2003:NuclearLaserSpectrosco_, Campbell2012:SingleIonNuclearClock_, Flambaum2025:NuclearClockBasedTh_}.  
The doped-crystals and CEMS set-ups provide a huge number of $^{229}$Th nuclei, with particle number densities that can exceed $10^{18}$~cm$^{-3}$. At the same time, such clocks can potentially be built very compact \cite{Zhang2024:229ThF4ThinFilmsSolid_,Elwell2025:LaserbasedConversionEl_,Morgan2025:SpinlessCrystalHighper_,Morgan2025:DesignNewThoriumNucle_}.
On the downside, perturbing electric and magnetic fields due to the lattice affect the  $^{229}$Th transition energy on a $10^{-8}$ level and can vary even within a single host crystal. Other solid state effects, changes of temperature or the continuous degradation of the crystal from the permanent $\alpha$-decays of $^{229}$Th and its daughter products ultimately limit precision, accuracy, and reproducibility of solid-state nuclear clocks \cite{Higgins2025:TemperatureSensitivity_,Ooi2026:FrequencyReproducibilit_}. 
In contrast, the conditions of single ions and ion crystals confined in Paul traps can be well controlled, facilitating an orders-of-magnitude higher level of accuracy and reproducibility. But even for the trapped-ion approach electronic many-body effects might limit the performance \cite{Beloy2023:TrapInducedAcZeemanSh_,Dzuba2023:EffectsElectronsNuclea_}. The low number of clock nuclei in a Paul trap imposes high demands on power and linewidth of vacuum ultraviolet (VUV) clock lasers. Despite recent technological advances in VUV lasers \cite{Zhang2022:TunableVUVFrequencyCo_,Schonberg2023:BelowthresholdHarmonic_,Lal2025:ContinuouswaveLaserSou_,Xiao2026:ContinuouswaveNarrowlin_} the excitation of $^{229}$Th nuclei in Paul traps has not been achieved, yet.

In this publication, we discuss an approach to thorium nuclear clocks that differs from the hitherto existing ones in several key aspects:
Foremost, we consider single $^{229}$Th$^{q+}$ ions in their highest charge states, i.e., $^{229}$Th$^{q+}$ with electrons only in the $K$- and $L$-shell or bare $^{229}$Th$^{90+}$. 
We evaluate the feasibility of such single-ion highly ionized $^{229}$Th$^{q+}$ clocks implemented by means of quantum logic spectroscopy (QLS) 
\cite{Schmidt2005:SpectroscopyUsingQuant_} in a linear Paul trap. Hereby, we focus on one-electron $^{229}$Th$^{89+}$ (H-like), and, the fully ionized nucleus, $^{229}$Th$^{90+}$.  
Such highly charged $^{229}$Th$^{q+}$ stands out with respect to resilience against external perturbations. For the bare nucleus electronic effects are absent, which makes  $^{229}$Th$^{90+}$ the most intriguing candidate for a \emph{nuclear} frequency standard. As has been discussed in 
\cite{Schiller2007:HydrogenlikeHighlyChar_, Yudin2014:MagneticDipoleTransitio_, Oreshkina2017:HyperfineSplittingSimp_, Kozlov2018:HighlyChargedIonsOpti_,Micke2020:CoherentLaserSpectrosc_,King2022:OpticalAtomicClockBas_,Gilles2024:QuadraticZeemanElectri_}, in highly charged ions such as few-electron $^{229}$Th$^{q+}$, owing to the very tight electron binding, shifts of clock transitions due to external fields are exceedingly suppressed compared to ions in low charge states.   
Furthermore, the electronic part of such simple quantum systems can be described by theory \emph{ab initio} on the level of full quantum electrodynamics (QED), and thus, the interpretation of results is not masked by many-body effects \cite{Kozhedub2008:NuclearDeformationEffe_, Ullmann2017:HighPrecisionHyperfine_,Skripnikov2018:NewNuclearMagneticMom_}. Due to a strong scaling of nuclear effects with the principal quantum number, nuclear-size contributions or hyperfine effects are strongly amplified and can be determined with small uncertainties. Related parameters such as nuclear charge radius or deformation as well as higher nuclear moments are important input data to deduce accurate estimates of the  $^{229}$Th sensitivity to variations of fundamental constants \cite{Yudin2014:MagneticDipoleTransitio_,Oreshkina2017:HyperfineSplittingSimp_,Flambaum2004:LimitsVariationsQuark_, Berengut2009:ProposedExperimentalMe_,Fadeev2020:Sensitivity$^229mathrmT_, Beeks2025:FinestructureConstantS_,Caputo2025:SensitivityNuclearCloc_}.

This universal detection method of QLS enables spectroscopy of all discussed charge states by mapping the nuclear excitation
on the qubit of a co-trapped logic ion. This qubit of the logic ion can be read-out with high fidelity. Thus, clock comparisons of different charge states as well as of different transitions in the $^{229}$Th ion can be carried out in the same set-up in order to control systematic influences or disentangle different physics effects. Such \emph{in situ} comparisons are particularly valuable for transitions with different sensitivities to fundamental physics \cite{Ludlow2015:OpticalAtomicClocks_,Kozlov2018:HighlyChargedIonsOpti_,Safronova2018:SearchNewPhysicsAtoms_,Shabaev2001:TestQEDInvestigations_}.

\begin{figure}[!tb]
\centering
\includegraphics[width=1.0\columnwidth,clip=true, trim = 0cm 0cm 1.5cm 0cm ]{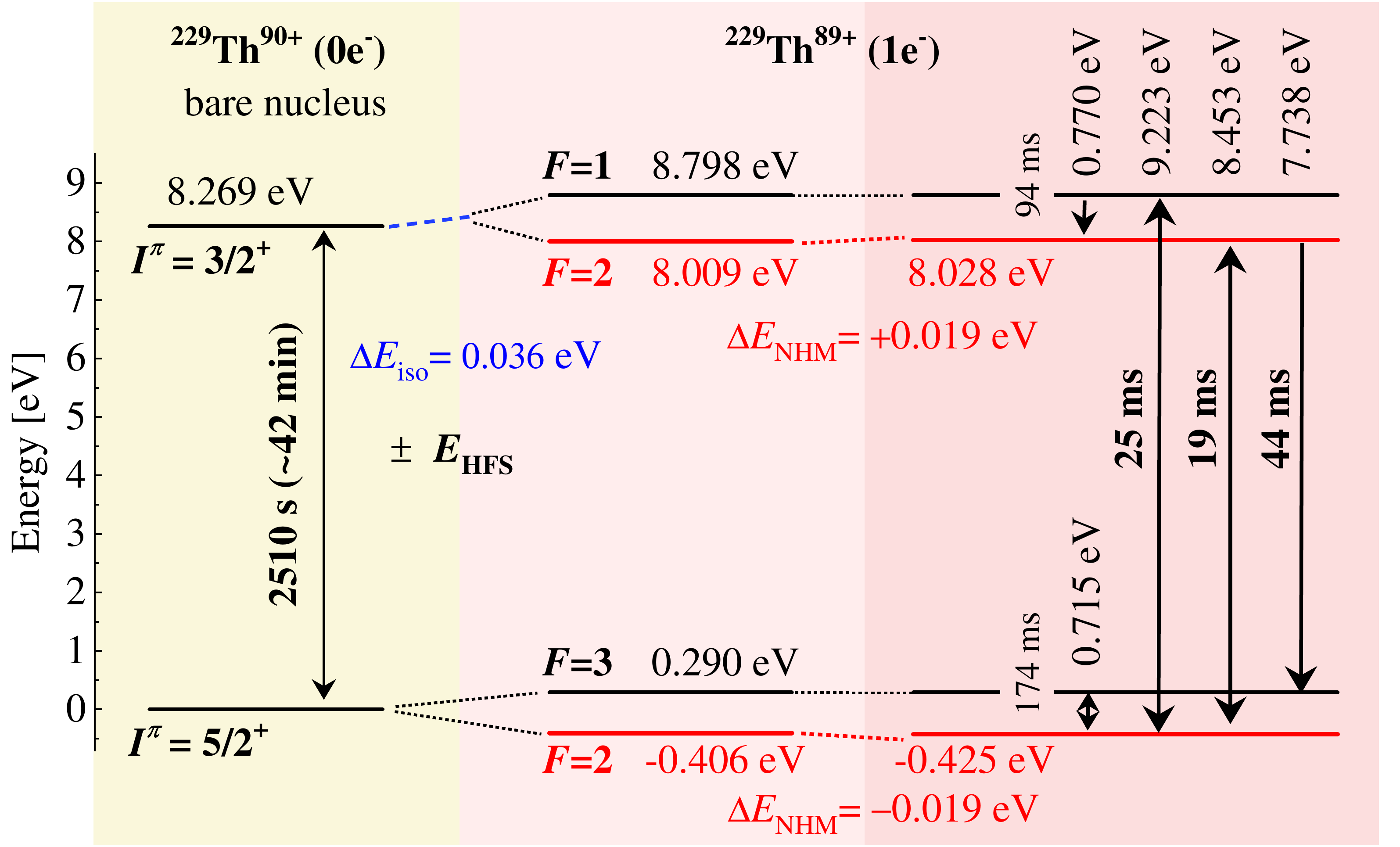}
\caption{\protect Nuclear clock transitions between the $I^{\pi} = 5/2^+$ ground state and the $3/2^+$ isomeric state for the bare nucleus, $^{229}$Th$^{90+}$ (left), and for H-like $^{229}$Th$^{89+}$ (center and right). 
For $^{229}$Th$^{89+}$ the density of the $1s$-electron shifts the energy by $\Delta E_\mathrm{iso} = 0.036$~eV. (Center) Ordinary HFS without NHM. (Right) HFS+NHM including $\Delta E_\mathrm{NHM} = \pm 0.019$~eV for the $F=2$ states (red). The VUV transitions between ground and isomeric state are vastly accelerated. For Li-like  $^{229}$Th$^{87+}$ the VUV transition energies are 8.494~eV, 8.363~eV and 8.248~eV and the corresponding lifetimes 900~ms, 549~ms and 982~ms.}
\label{fig:229Th_quenching}
\end{figure}

A decisive asset using highly ionized $^{229}$Th$^{q+}$ is the process of Nuclear Hyperfine Spin Mixing (NHM) 
\cite{Lyuboshitz1966:_, Wycech1993:PredictionsNuclearSpin_,Karpeshin1998:RatesTransitionsHyperf_, Pachucki2001:NuclearspinMixingOscil_, Tkalya2016:MagneticHyperfineStruc_, Shabaev2022:GroundState$g$FactorH_} which can be utilized to `tune' the natural linewidth of the nuclear-clock resonance over more than five orders of magnitude. In few-electron $^{229}$Th$^{q+}$ unpaired electrons induce strong magnetic fields at the site of the nucleus leading to magnetic hyperfine interactions (HFI). Beyond the ordinary hyperfine splitting (HFS), the HFI mixes states with the same total angular momentum $F=2$ of nuclear ground $(g)$ and isomeric state $(m)$ and leads to small energy shifts $\Delta E_\mathrm{NHM}$ for the $F=2$ states. Most notably, the lifetime of the $^{229}$Th isomeric state is drastically shortened, in H-like $^{229}$Th$^{q+}$ by more than five orders of magnitude \cite{Wycech1993:PredictionsNuclearSpin_,Karpeshin1998:RatesTransitionsHyperf_, Pachucki2001:NuclearspinMixingOscil_, Tkalya2016:MagneticHyperfineStruc_, Shabaev2022:GroundState$g$FactorH_}.  The effect depends on the size of the magnetic field, hence, NHM decreases for Li-like or B-like ions and for higher atomic shells. NHM is absent for paired electrons such as in He-like or Be-like ions. NHM is thus a unique tool towards a single-ion no-electron nuclear clock since lifetimes of several ten milliseconds in $^{229}$Th$^{89+}$, about one second in $^{229}$Th$^{87+}$ or one to two minutes in $^{229}$Th$^{85+}$ (Fig.\ \ref{fig:229Th_quenching} and \cite{Shabaev2022:GroundState$g$FactorH_} and Supplementary Material therein)  ease the requirements for narrow-band VUV-lasers and speed up development and commissioning of the experiment. For comparison, the radiative vacuum lifetime of the bare nucleus $^{229}$Th$^{90+}$ or of lowly charged $^{229}$Th$^{q+}$ ions is about $\tau_\gamma \sim 42$~min \cite{Tiedau2024:LaserExcitationTh229N_}.
The effect of NHM is visualized in Fig.\ \ref{fig:229Th_quenching} for H-like $^{229}$Th$^{89+}$. The left part shows the transition energy $E_\gamma^{(m, 90+)}$ and the vacuum lifetime \cite{Tiedau2024:LaserExcitationTh229N_} for fully ionized $^{229}$Th$^{90+}$. $I$ and $\pi$ are nuclear spin and parity, respectively. The different electron densities at the site of the nucleus \cite{Dzuba2023:EffectsElectronsNuclea_, Perera2025:HostDependentFrequency_,Zheng2026:EnhancedSensitivityNuc_} result in an isomeric energy for the bare nucleus $E_\gamma^{(m, 90+)}$ that we calculated to be lower by $\Delta E_\mathrm{iso} = 0.086$~eV compared to the one of the $^{229}$Th:CaF$_2$-crystal $E_\gamma^{(m, 4+)} \approx 8.3557$~eV. This shift $\Delta E_\mathrm{iso}$ is analogous to the so-called isomer or chemical shift in Mössbauer spectroscopy. In our calculations for $\Delta E_\mathrm{iso}$ we assume a difference in the mean square charge radii of $\delta \langle r^2 \rangle =  \langle r^2_{229m}\rangle - \langle r^2_{229}\rangle = 0.0103\, \mathrm{fm}^2$  corresponding to a weighted mean of \cite{Yamaguchi2024:LaserSpectroscopyTripl_} and \cite{Safronova2018:NuclearChargeRadii$^2_}. Our computations were performed using the MDFGME multi-configuration Dirac-Fock code \cite{:MDFGME_}. The nuclear charge distribution is modelled using a 2-parameter Fermi distribution. Taking into account the different value for $\delta \langle r^2 \rangle$, our calculated  $\Delta E_\mathrm{iso}$ data are in agreement with the ones in \cite{Perera2025:HostDependentFrequency_,Zheng2026:EnhancedSensitivityNuc_}. 

In H-like $^{229}$Th$^{89+}$, the $1s$-electron increases $E_\gamma$ by $\Delta E_\mathrm{iso} = 0.036$~eV with respect to the bare nucleus and enables the HFI. In the center of Fig.\ \ref{fig:229Th_quenching}, only the ordinary HFS is displayed while in the right part the additional influence of mixing is included. NHM slightly repels the two $F=2$ levels by $\Delta E_\mathrm{NHM} = \pm 0.019$~eV and accelerates the isomeric transition $\sim 10^5$-fold. Our re-evaluation for HFS and NHM are performed based on \cite{Shabaev2022:GroundState$g$FactorH_} employing the most recent values for the magnetic moments $\mu^{(g)} = 0.366(6)  \mu_\textrm{N}$  \cite{Porsev2021:PrecisionCalculationHy_} for the ground state and $\mu^{(m)} =  -0.378(8) \mu_\textrm{N}$ \cite{Yamaguchi2024:LaserSpectroscopyTripl_} for the isomeric state as well as a reduced nuclear transition probability for the $M1$ $\gamma$-transition of $B(M1) = 0.022$~Weisskopf units.~\cite{Tiedau2024:LaserExcitationTh229N_}. For H-like, Li-like and B-like $^{229}$Th$^{q+}$, HFS, NHM lifetimes and energies for different $\mu_\textrm{N}$ or $B(M1)$  values can be obtained using the formulas and/or tables in \cite{Shabaev2022:GroundState$g$FactorH_} and its supplementary material. Up to now, due to the high sensitivity on the one hand and the experimental uncertainties for $\delta \langle r^2 \rangle$, the nuclear magnetic moments $\mu^{(g, m)}$ and $B(M1)$ on the other hand, the transition energies for highly ionized $^{229}$Th$^{q+}$ still possess comparatively large uncertainties. 

Highly charged $^{229}$Th$^{q+}$ ions with unpaired $j=1/2$ valence electron are unique laboratories for testing fundamental physics since they possess five clock transitions, three of which are accessible with lasers from the $F=2$ ground state: The two VUV transitions to either the $F=2$ level ($2_g \to 2_m$) or the $F=1$ level ($2_g \to 1_m$)  of the isomeric state, and the $2_g \to 3_g$ ground-state HF transition. The difference of the two VUV transitions yields the HF-splitting of the isomeric state while the difference of $2_g \to 2_m$ and the ground state HF transition yields the 5th VUV transition $3_g \to 2_m$. The HF-splitting including the NHM contribution for the nuclear ground state is $\Delta E_\mathrm{HFS} = 0.715$~eV for H-like  $^{229}$Th$^{89+}$ and 0.115~eV for the Li-like ion, both accessible by infrared (IR) lasers. For the isomer $\Delta E_\mathrm{HFS}$ = 0.770~eV for $^{229}$Th$^{89+}$ and $\Delta  E_\mathrm{HFS}$ = 0.131~eV for $^{229}$Th$^{87+}$. The different excitation options enable comparisons of nuclear and atomic clock transitions in the same ion. The individual transitions feature very different sensitivities to potential variations of $\alpha$, $m_e/m_p $, and $X_q$. For instance, using equation 3 in \cite{Oreshkina2017:HyperfineSplittingSimp_} yields $K_\alpha \approx 6$ for the atomic HF transition in H-like $^{229}$Th$^{89+}$ whereas $K_\alpha \approx 6000 $ \cite{Beeks2025:FinestructureConstantS_} for the nuclear transition. In addition, a comparison of the HF transitions of ground and isomeric state allows one to isolate potential variations of the strong interaction since the sensitivities for variations of $m_e/m_p $ and $\alpha$ are identical.
Up to now, no estimate for the sensitivity of the contribution $\Delta E_\mathrm{NHM}$ due to NHM is available. However, in $^{229}$Th$^{87+}$ $\Delta E_\mathrm{NHM}$ is already reduced by a factor of $\sim$ 3900 compared to $^{229}$Th$^{89+}$. As is pointed out, e.g., in \cite{Shabaev2001:TestQEDInvestigations_,Kozhedub2008:NuclearDeformationEffe_,Zheng2026:SimultaneousDeterminati_} the comparison of different charge states of highly charged ions with $j=1/2$ allows for a clear-cut interpretation with respect to nuclear parameters such as $\delta \langle r^2 \rangle $, magnetic dipole moments, or nuclear deformation since wanted and unwanted contributions to the transition energies can be disentangled.
Furthermore, a comparison of the bare $^{229}$Th$^{90+}$ with $^{229}$Th$^{q+}$ with paired electrons, such as He-like $^{229}$Th$^{88+}$ ($\Delta E_\mathrm{iso}  = 0.070$~eV) provides an unbiased and highly significant access to the nuclear charge distribution \cite{Perera2025:HostDependentFrequency_, Zheng2026:EnhancedSensitivityNuc_}. The different radial overlap of the electron densities of the $2s^2_{1/2}$ and $2p^2_{1/2}$ electrons in Be-like $^{229}$Th$^{86+}$ ($\Delta E_\mathrm{iso}  = 0.080$~eV) and C-like $^{229}$Th$^{84+}$ ($\Delta E_\mathrm{iso}  = 0.084$~eV) would even enable one to probe different parts of the nucleus and thus derive deformation information \cite{Kozhedub2008:NuclearDeformationEffe_}. 

The implementation of a highly-charged-$^{229}$Th$^{q+}$ single-ion clock is feasible through QLS in a linear Paul trap. Such a QLS set-up is presently being realized at the Highly Charged Ion Trapping facility (HITRAP)\cite{Kluge2008:HITRAPFacilityGSIHigh_, Rausch2026:DecelerationAccelerator_} at the GSI Helmholtzzentrum für Schwerionenforschung in Darmstadt, Germany. The highly charged $^{229}$Th$^{q+}$ ions are artificially synthesized in-flight at high energies and are subsequently separated  similar to the method described in \cite{Brandau2009:IsotopeShiftsDielectro_,Brandau2013:ProbingNuclearProperti_}. The  beam of  $^{229}$Th$^{q+}$ ions is decelerated, first in the heavy ion storage ring ESR, and finally brought to rest with the HITRAP facility \cite{Rausch2026:DecelerationAccelerator_}. From the HITRAP cooler trap  single $^{229}$Th$^{q+}$ ions can be transferred into a linear Paul trap for QLS. Sufficient ion storage times of months even for the extreme charge states up to $^{229}$Th$^{90+}$ can be achieved through trap operation inside a cryogenic enclosure at 4~K  \cite{Diederich1998:ObservingSingleHydroge_, Diederich1999:GfactorHydrogenlikeIon_, Ulmer2026:AntiprotonTrapping614_}. For the single-ion $^{229}$Th$^{q+}$ QL clock a second ion species is required for sympathetic cooling and state detection.  Owing to the high charge-to-mass ratio of $^{229}$Th$^{q+}$, $^9$Be$^+$ is an excellent candidate. Accordingly, a two-ion crystal composed of one $^{229}$Th$^{q+}$ and one $^9$Be$^+$ ion is prepared along the symmetry axis of the linear Paul trap at the radio-frequency null. Laser cooling down to the quantum-mechanical ground state of motion \cite{Monroe1995:ResolvedSidebandRamanC_} for at least one motional mode enables QL operations \cite{Schmidt2005:SpectroscopyUsingQuant_}. Thereby, the nuclear or atomic hyperfine state can be mapped on the qubit of the $^9$Be$^+$ ion through a series of laser pulses. Finally, the qubit can be read-out with high fidelity through resonant scattering of the light of a detection laser on a fast $E1$ transition of the logic ion.

We evaluate the timescales and feasibility of QLS, and represent, for simplicity, the relevant transition in the $^{229}$Th$^{q+}$ ion by a two-level system. We consider either a VUV frequency comb or a continuous-wave (cw) laser. While a cw laser is a simpler clock laser system with the benefit of only exposing a single frequency and a lower total optical power to the ions, the broad frequency-comb pattern is advantageous for the initial search of the transition \cite{Zhang2024:FrequencyRatio229mThN_}. 
The excitation process is described by the optical Bloch equations with relaxation terms to account for spontaneous emission at a decay rate $\Gamma = 1/\tau$. Here, $\tau$ is the natural lifetime of the excited state. The laser has a linearly polarized light field and a FWHM-linewidth $\Gamma_L$ in angular units. The coherence of the excitation process degrades with 
$    \tilde{\Gamma} = (\Gamma + \Gamma_L) /2 $ \cite{vonderWense2020:TheoryDirectLaserExci_, vonderWense2020:ConceptsDirectFrequenc_}. 
For the Rabi frequency $ \Omega = 2 | \langle e | \hat{H}_I |g \rangle | / \hbar $  one obtains
\begin{equation}
    \Omega = \sqrt{\frac{2 \pi c^2 I_L C_{ge}^2 G^2 \, \Gamma}{\hbar \omega_0^3}}
\end{equation}
where $I_L=2 c \varepsilon_0 E_0^2$ is the laser light intensity, $C_{ge}$ the Clebsch-Gordan coefficients, $G$ a geometric factor accounting for the laser polarization, angle of incidence and the specific Zeeman component. $\omega_0$ is the transition frequency. According to Torrey's solution \cite{Noh2010:AnalyticSolutionsOptic_,vonderWense2020:TheoryDirectLaserExci_} to the optical Bloch equations, the on-resonant (i.e. $\omega_L = \omega_0$) time-dependent excitation probability is then given by
\begin{equation}
    P(t) = \frac{\Omega^2}{2\left( \Gamma \, \tilde{\Gamma} + \Omega^2 \right)} \left[ 1 - e^{-\Gamma^\prime t} \left( \cos{\left( \lambda t \right) } + \frac{\Gamma^\prime}{\lambda} \sin{\left( \lambda t \right) }  \right) \right]
    \label{eq:ExProb}
\end{equation}
with $\Gamma^\prime = (\Gamma + \tilde{\Gamma})/2$; 
$    \lambda = \sqrt{|\Omega^2-( \tilde{\Gamma} - \Gamma )^2 / 4 |} $
and $|\tilde{\Gamma}-\Gamma|/2 < \Omega$. 
\begin{figure}[tb]
\centering
\includegraphics[width=0.9\columnwidth, clip=true, trim = 0.cm 0.cm 0.cm 0.cm ]{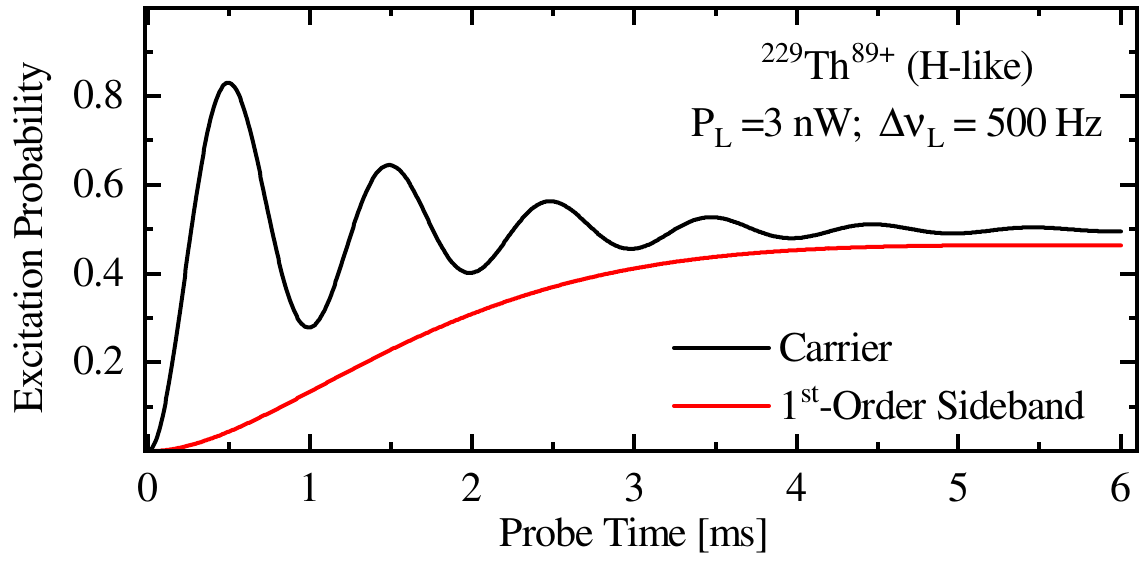}
\includegraphics[width=0.9\columnwidth, clip=true, trim = 0.cm 0.cm 0.cm 0.cm ]{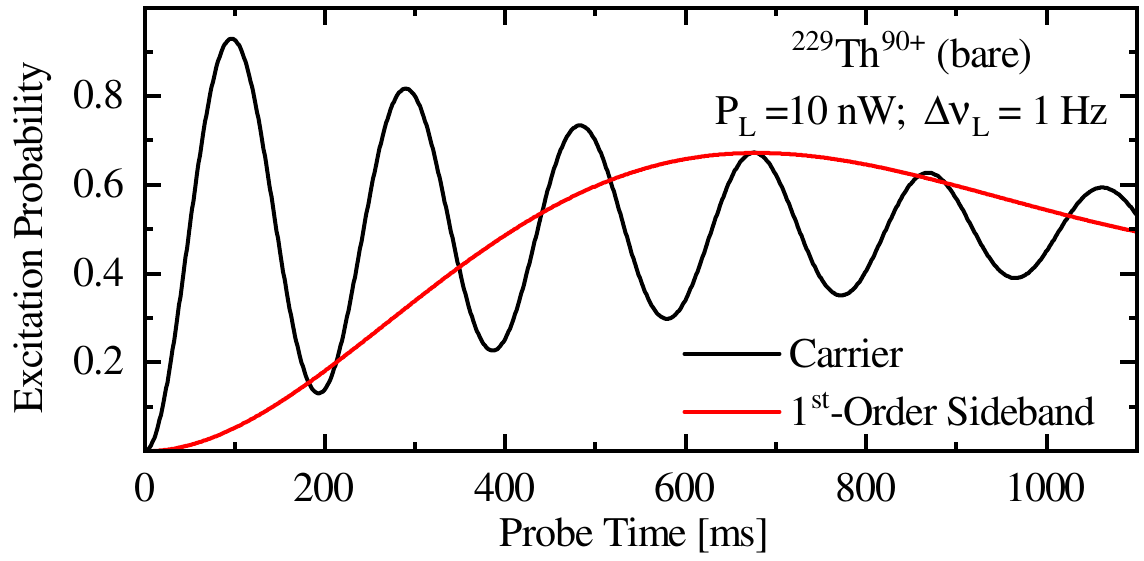}

\caption{\protect (Top) Excitation probability of the isomer for the $2_g \to 2_m$ transition in H-like $^{229}$Th$^{89+}$ with laser power $P_L = 3$~nW and laser linewidth (FWHM) $\Gamma_L = 2\pi \cdot 500$~Hz corresponding to a comb tooth of a stabilized VUV frequency comb. (Bottom) Excitation probability for the $3/2^+  \to 5/2^+$ nuclear transition in bare $^{229}$Th$^{90+}$ with $P_L = 10$~nW and $\Gamma_L = 2\pi \cdot 1$~Hz. Carrier and first-order side-band excitation are displayed in black and red. }
\label{fig:Excitation_probability}
\end{figure}
QL algorithms also involve coherent (de)excitation on motional sidebands, to couple the internal nuclear and external motional states. For the first-order sidebands, the excitation frequency is $\omega_L = \omega_0 \pm \omega_u$, where $\omega_u$ is the frequency of a specific motional mode of the two-ion crystal in the harmonic potential of the Paul trap. The Rabi frequency of the first-order sidebands is suppressed by the Lamb-Dicke parameter $\eta$. Typically $\eta \ll1$ to facilitate quantum control and ground-state cooling of the motional modes. In the following we assume an axial out-of-phase motional frequency of $\omega_u = 2\pi \cdot 1.3$~MHz and an axial out-of-phase component of the normalized motional mode eigenvector \cite{Kielpinski2000:SympatheticCoolingTrap_,Wubbena2012:SympatheticCoolingMixe_} of about 0.86 for a $^9$Be$^+$-$^{229}$Th$^{q+} (q=89, 90)$ two-ion-crystal, cf. \cite{Kozlov2018:HighlyChargedIonsOpti_}. 
Excitation with the VUV laser beam along the axial symmetry axis of the linear Paul trap yields an axial out-of-phase Lamb-Dicke parameter $\eta \approx 0.15$  for $^{229}$Th$^{89+}$ and $^{229}$Th$^{90+}$. In order to calculate $\Omega$ and $P(t)$, a 5 $\mu$m-diameter laser focus is used. $C_{ge}^2 G^2$ is set to 0.5 for the nuclear excitation between the $2_g \to 2_m$  hyperfine levels in $^{229}$Th$^{89+}$ or the  $3/2^+  \to 5/2^+$ levels in $^{229}$Th$^{90+}$.
For H-like $^{229}$Th$^{89+}$ we adopt a VUV frequency comb with a comb tooth power $P_L = 3$~nW and a FWHM linewidth  $\Gamma_L = 2\pi \cdot 500$~Hz while for the bare nucleus $^{229}$Th$^{90+}$ we assume a highly-stable VUV cw laser with $P_L = 10$~nW and a much narrower linewidth of $\Gamma_L = 2\pi \cdot 1$~Hz. Laser powers up to 100~nW at a linewidth of less than 100~Hz for a VUV cw laser have recently been demonstrated \cite{Xiao2026:ContinuouswaveNarrowlin_}. Using the same technique, powers might even reach 100~$\mu$W and linewidths of $<1$~Hz in the near future.  Fig.\ \ref{fig:Excitation_probability} shows the expected coherent Rabi flopping on the nuclear transition via the carrier and first-order sideband and demonstrate the feasibility of QLS for H-like $^{229}$Th$^{89+}$ and the bare nucleus $^{229}$Th$^{90+}$.\\

In summary, we have discussed $^{229}$Th$^{q+}$  nuclear clocks with ions in their highest ionization stages. Such clocks possess a set of features that distinguishes them from their counterparts in low charge states due to the simplicity of the quantum systems. Thus, highly ionized $^{229}$Th$^{q+}$ has a huge potential for understanding the $^{229}$Th nucleus and also possesses a long list of applications in fundamental physics. The tunability of the clock-linewidth through a deliberate choice of the charge state exploiting NHM is a decisive asset towards the development of a `pure' nuclear clock, namely a clock consisting of only the nucleus itself. 
Beyond the production and investigation of highly ionized $^{229}$Th$^{q+}$ at heavy-ion accelerators, a dissemination of the technology to laboratories world-wide using high-performance electron beam ion traps (EBIT) seems feasible \cite{Marrs1994:ProductionTrappingHydr_,Elliott1995:WireProbeIonSource_,Elliott1996:TrappedIonTechniqueMea_,Beiersdorfer1997:MeasurementsNuclearPar_,Hu2012:ExperimentalDemonstrati_}.
 The very long storage times  that can be achieved in a 4~K environment would even allow for the distribution of the highly charged $^{229}$Th$^{q+}$ ions from dedicated production sites such as the GSI or dedicated `$^{229}$Th$^{q+}$-EBIT factories' to remote metrology laboratories with transportable Penning traps as has recently been demonstrated with antiprotons at CERN \cite{Leonhardt2025:ProtonTransportAntimat_,Smorra2026:RoadTransportTrappedA_}.\\

This project has received funding from the European Research Council (ERC) under the
European Union’s Horizon 2020 research and innovation programme (Grant agreement No. 101142155). P.I. is a member of the Helmholtz Alliance HA216/EMMI. P.I. thanks the Institute of Physics of CNRS for financing the high-performance computer used to perform calculations for this work. P.M. acknowledges funding by the Initiative and Networking Fund of the Helmholtz Association under the call Helmholtz-Nachwuchsgruppen (VH-NG-19-23).\\

\textbf{End Matter}\\
In the majority of existing experimental concepts $^{229}$Th is either obtained from $\alpha$-decay of $^{233}$U, or alternatively, minute quantities of the precious $^{229}$Th are directly used to dope the clock crystals or produce thin film samples. This is in part due the fact that worldwide an estimated amount of less than 40~g of sufficiently isotopically clean $^{229}$Th exists \cite{Beeks2021:Thorium229LowenergyIso_}. In our approach we use a distinctively different method to obtain the required $^{229}$Th$^{q+}$ ions. At the accelerator facilities of GSI we have produced and separated sufficient amounts of highly charged $^{229}$Th$^{q+}$ ions in-flight using a relativistic $^{238}$U beam with velocities of 80\% to 90\% speed-of-light that impinges onto a thick production target. Up to now, a 1~cm beryllium metal plate and a combination of 2.4~cm aluminium with an additional thin copper foil have been tested as targets. In the collision, a cocktail of highly ionized radioisotopes including the desired $^{229}$Th$^{q+}$ is synthesized in-flight \cite{Brandau2009:IsotopeShiftsDielectro_,Steck2010:NovelMethodPreparation_, Brandau2010:ResonantRecombinationI_, Brandau2013:ProbingNuclearProperti_} and is down-beam injected into the heavy ion storage ring ESR. There, the ions are phase-space cooled by means of stochastic and electron cooling \cite{Steck2004:ElectronCoolingExperim_}. Subsequently, the $^{229}$Th$^{q+}$ ions are separated from the primary beam and other unwanted isotopes using the ring itself as a separator. This separation approach is well-established \cite{Brandau2009:IsotopeShiftsDielectro_,Steck2010:NovelMethodPreparation_, Brandau2010:ResonantRecombinationI_, Brandau2013:ProbingNuclearProperti_}, is highly selective and leads to a very clean and intense secondary beams. By accumulation of several injections \cite{Nolden2013:RadioactiveBeamAccumul_,Glorius2023:StorageAccumulationDec_,Leckenby2024:Hightemperature205TlDe_}intensities close to $10^5$ $^{229}$Th$^{89+}$ ions could already be obtained with the potential to increase this number by an order of magnitude  \cite{Glorius2023:StorageAccumulationDec_,Leckenby2024:Hightemperature205TlDe_}. The ESR also serves as a deceleration device from relativistic energies down to 4 MeV/u. In the initial experiments on the production of $^{229}$Th$^{89+}$ a first deceleration step down to 190~MeV/u could be shown.  After extraction from the ring at 4 MeV/u, further deceleration will then be achieved with the HITRAP decelerator. The slowed-down ions are subsequently captured in a cooler trap \cite{Rausch2022:CommissioningHITRAPCoo_} and further distributed to individual experiments such as Penning and Paul traps for precision experiments with ions at rest. Currently, a  $^{232}$Th ion source is being developed for the GSI injector. With a $^{232}$Th  primary beam significantly higher yields of the secondary ions $^{229}$Th$^{q+}$ can be assumed in the future due to the much higher nuclear cross section $^{232}$Th$\to$$^{229}$Th compared to $^{238}$U$\to$$^{229}$Th. 
High nuclear production yields for $^{229}$Th$^{q+}$ ions are primarily only achieved for ions with 0 to 2 electrons. Yet, in the storage ring itself \cite{Brandau2009:IsotopeShiftsDielectro_} or in one of the ion traps of HITRAP essentially any desired charge state of $^{229}$Th can be obtained employing atomic charge-exchange processes. 
It is noted that the in-flight production leads to an initial yield of about 50\% $^{229}$Th$^{q+}$ ions already in the isomeric state \cite{Karpeshin1998:RatesTransitionsHyperf_}. 


%

\end{document}